\documentclass[preprint,aps,prl]{revtex4}
\usepackage{times}
\usepackage{amsmath,amscd,amssymb,latexsym}
\usepackage{graphicx}
\usepackage{epstopdf}

\newcommand{\ee}{\end{eqnarray}}
\newcommand {\be}[1]{\begin{eqnarray} \mbox{$\label{#1}$}  }

\newcommand{\diff}{\mathrm{d}}

\newcommand{\partdiff}[2]{\frac{\partial #1}{\partial #2}}

\newcommand{\ket}[1]{\left | #1 \right \rangle}
\newcommand{\bra}[1]{\left \langle #1 \right |}

\newcommand{\mean}[1]{\left \langle #1 \right \rangle}

\newcommand{\com}[2]{\left[ #1, #2 \right]}

\newcommand\ie {{\it i.e.}, }

\newcommand{\nn}{\nonumber\\}
\newcommand{\noi}{\noindent}

\newcommand\suml{\sum\limits }

\newcommand\prodl{\prod\limits }

\newcommand\half{\frac 1 2 }
\newcommand{\pref}[1]{(\ref{#1})}

\renewcommand{\nn}{\nonumber\\}
\renewcommand{\noi}{\noindent\\}

\newcommand{\pa}[1]{\partial_{#1}}

\newcommand\e {\epsilon}

\newcommand\ra{\rightarrow}
\newcommand\bs{\boldsymbol}

\begin{document}

\title{The Fulling-Unruh effect in general stationary accelerated frames}
\author{Jan Ivar Korsbakken$^{a,b}$, and Jon Magne Leinaas$^{a,b}$}
\affiliation{${(a)}$Department of
Physics,University of Oslo, P.O. Box 1048 Blindern, 0316 Oslo, Norway}
\affiliation{${(b)}$Department of Physics, University of California
at Berkeley, Berkeley, CA 94720, USA}

\date{May 30, 2004}

\begin{abstract}
We study the generalized Unruh effect for accelerated reference 
frames that include
rotation in addition to acceleration. We focus particularly on the 
case where the motion
is planar, with presence of a static limit in addition to the event 
horizon. Possible
definitions of an accelerated vacuum state are examined and the 
interpretation of the
Minkowski vacuum state as a thermodynamic state is discussed. Such a
thermodynamic state is shown to depend on two parameters, the
acceleration temperature and a drift velocity, which are determined 
by the acceleration
and angular velocity of the accelerated frame. We relate the 
properties of Minkowski
vacuum in the accelerated frame to the excitation spectrum of a 
detector that is
stationary in this frame.  The detector can be excited both by 
absorbing positive energy
quanta in the "hot" vacuum state and by emitting negative energy 
quanta into the
"ergosphere" between the horizon and the static limit. The effects 
are related to similar
effects in the gravitational field of a rotating black hole.
\end{abstract}

\pacs{PACS: }
\maketitle
\section{Introduction}
A radiation detector that is uniformly accelerated through vacuum will be
excited as if vacuum were hot,
with a temperature
proportional to the acceleration \cite{ref:UnruhBlackHoleEvaporation}. The effect is small, but from a theoretical point of view it is important and relates to interesting questions concerning
the definition of vacuum states and particle excitations in curved
spacetime \cite{Fulling73},
\cite{Davies75}. Indeed,
the effect is closely related to the Hawking effect
\cite{ref:HawkingBlackHoleRadiation1},
\cite{ref:HawkingBlackHoleRadiation2} and is equivalent to the
temperature effect measured by a stationary detector close to the
event horizon of a black hole, in the limit where the mass of the
black hole tends to infinity \cite{ref:UnruhBlackHoleEvaporation}.

Uniform linear acceleration corresponds to motion along a
hyperbolic spacetime curve. A detector moving along this curve will
experience a time-independent situation and hence settle in a
stationary state. However, there exist also other time-like curves with
the same property, where the spectrum of
vacuum fluctuations is independent of the detector's proper time
parameter and where the detector therefore will settle into a stationary
state with a non-trivial distribution over excited states \cite{ref:LetawPfautschStationaryQuantizedField},
\cite{ref:LetawStationaryDetectorExcitation}. In general these curves
involve rotation in addition to acceleration.

In the case of motion along stationary curves other than those with
uniform linear acceleration, the excitation spectrum of the detector
will not have a truly thermal form
\cite{ref:LetawStationaryDetectorExcitation}. However, even with
rotation involved, for a two-level system an effective temperature may
be defined which depends only weakly on the energy splitting and on
detector-dependent variables. The question of a ``circular Unruh
effect'' has been discussed in particular with relation to
polarization effects of electrons in storage rings
\cite{ref:BellLeinaas1}, \cite{ref:BellLeinaas2}, \cite{Unruh99}, \cite{Leinaas01}. Also  for electrons  circulating in a cavity the circular Unruh effect has been considered \cite{Rogers88}, \cite{Levin88}, \cite{Davies96}.

The excitation spectrum of a detector coupled to a scalar field
was examined several years ago by
Letaw \cite{ref:LetawStationaryDetectorExcitation} for all types
of stationary curves, and the possibility of defining more general
``accelerated vacua'' and particle states for observers travelling along
all possible stationary curves was subsequently discussed in an
interesting paper by Letaw
and Pfautsch \cite{ref:LetawPfautschStationaryQuantizedField}. Their
conclusion was that only two distinct inequivalent vacuum states can be
defined in flat spacetime, termed the Minkowski vacuum and the Fulling
vacuum, where the former seems to contain a thermal spectrum of particles
relative to the latter. They point out however, that this definition does
not seem to agree with the excitation spectra of stationary particle
detectors whenever rotation is involved, since a particle detector on a
trajectory involving rotation will become excited even when the  vacuum state associated with that trajectory is the Minkowski vacuum, and in general it will
not exhibit a purely thermal excitation spectrum for trajectories corresponding to the Fulling vacuum state.

In this paper we follow up and extend the discussion of Letaw and 
Pfautsch on the
generalization of the Unruh effect to general stationary timelike
curves. We first focus on the definition of the Hamiltonian for a quantum
system in the accelerated frame defined by a general stationary 
space-time curve. As
a specific case we consider a free Klein Gordon field. Next we 
reconsider the case of
linear acceleration where we focus on the Boguliubov transformation 
which relates
the Minkowski vacuum state to the Fulling vacuum. The central section of the
paper contains a discussion of the causal structure of space-time as 
viewed in an
accelerated frame with rotation, and further, the implication of this for the
definition of vacuum states and the interpretation of the Minkowski vacuum as a
thermodynamic state.

In general the Hamiltonian, defined as the time evolution operator in 
the accelerated
frame, is not bounded from below and the standard definition of the 
vacuum as the
ground state of the Hamiltonian is therefore not applicable. In a more general
definition, the vacuum is an eigenstate of the Hamiltonian, which 
may allow radiation
quanta with both positive and negative energy. The Minkowski vacuum 
state is, in
this sense, a vacuum state for all the stationary accelerated frames, and the
excitation of an accelerated detector, in the same  formulation, is due to the
emission of negative energy quanta. However, for accelerated frames 
which possess an
event horizon, alternative definitions are possible, due to a 
symmetry in the energy
spectrum related to PCT invariance \cite{Sewell82,Hughes85,Bell85}. A 
specific case is
given by the Fulling vacuum state which is based on a separation of 
the field modes
associated with the two sides of the horizon. For linear acceleration 
this separation
will also push the negative energy excitations behind the event 
horizon and the Fulling
vacuum becomes the true ground state of the system restricted to one 
side of the horizon.
For other types of motion this is not the case. The field modes 
associated with the
causally disconnected region behind the horizon can be separated from the field
modes of the ``physical" region, but the presence of excitations with 
negative energy
is no longer avoided. This affects the excitation spectrum of an accelerated
detector, which can gain energy by emitting negative energy quanta as well as
absorbing positive energy quanta.

The discussion of vacuum states and excitations in the accelerated frames is
supplemented by a calculation of the effective temperature, for 
various degrees of
rotation. It is stressed that whereas the definition of non-trivial 
vacuum states and
interpretation of the Minkowski vacuum as a thermodynamic state 
depends on the asymptotic
properties of the stationary space-time curves (through the existence 
of an event
horizon), this is not so for the excitation rates of a detector, 
which are determined
only by a limited part of the space-time curve. This explains why the 
excitation spectrum
changes smoothly with rotation even when the angular velocity exceeds 
the proper
acceleration and the event horizon disappears.

Although most of the discussion in the paper is focused on the case of planar
motion, we include a brief section where we examine the general case where the
space-time curve cannot be restricted to two space dimensions. In 
this case there is
always an event horizon, and most of the general discussions of 
planar curves with
horizons also apply to these cases.

The discussion of detector response for space-time curves with rotation can be
related to local effects in the gravitational field of a
rotating black hole. In the same way as the linearly accelerated 
frame can be viewed
as a stationary frame close to the horizon of a non-rotating black 
hole in the limit
where the black hole mass tends to infinity, the accelerated frames 
with rotation
can be viewed as limiting cases of stationary frames close to a 
rotating black hole
with large mass. In the Appendix we demonstrate this by showing how 
the flat space
metric of the accelerated frame is recovered from the Kerr metric in 
the limit where
the mass tends to infinity.

We shall in this paper use natural units, with $\hbar = c= 1$.

\section{Reference frames and Hamiltonians}
\label{sect:ReferenceFramesAndHamiltonians}
A time-like space-time curve $C$ defines a natural ``accelerated
coordinate system" in the following way. At an arbitrary point on the
curve an orthonormal reference frame is defined with one of the unit
vectors pointing along the trajectory. This vector defines the local time
axis. The local frame is transported to any other point on the curve by a
``Fermi-Walker transport", and each of these frames can further be
extended in the space-like direction to form a full rest frame at the
given (proper) time. Clearly such a coordinate system may have
(coordinate) singularities, but for a well behaved space time curve
there will always be a finite region around the curve without
singularities.

Let us consider a given space-time curve with its associated
accelerated (curvilinear) coordinate system. A quantum system, and
particularly a quantum field, which in the standard Minkowski space
formulation is described by a Hamiltonian $H$, will in the accelerated
reference frame be described, in a natural way, by a time evolution
operator of the form
\be{accham}
{H}_{O} \, = \, {H} - \bs a\cdot \bs {K} - \bs\omega\cdot \bs{J}
\label{eq:ObserverHamiltonian}
\ee
where $\bs{K}$ is the boost operator and $\bs{J}$ is the generator
of rotation in the Minkowski space description of the system. This form
for the time evolution operator follows when the quantum state at a given
proper time of the accelerated frame is identified with the quantum state of
the corresponding inertial rest frame \cite{ref:BellLeinaas1,Bell85}. The
Hamiltonian generates transformations between inertial rest frames at
different times and this transformation will, due to the acceleration,
involve Lorentz transformations in addition to time translation. By use of
the freedom of choice for the definition of three space directions
$(x,y,z)$ the Hamiltonian can further be brought into the form
\be{accham2}
{H}_{O} \, = \, {H} - a{K}_x - \omega_z {J}_z - \omega_x
{J}_x
\ee
which is the general form we shall apply for the time evolution operator
in the accelerated frame. As a specific example, we shall
consider the case of a real Klein-Gordon field $\phi$. The operators
then are given by
\be{KG}
{H}_{O}&=&\int d^3x \half(\pi^2+(\bs\nabla \phi)^2+m^2\phi^2) \nn
{\bs J}&=& \int d^3x \;\bs x \times \pi \bs\nabla \phi \nn
{\bs K}&=&-\int d^3x \;\half \bs x \,(\pi^2+(\bs\nabla \phi)^2+m^2\phi^2)
\ee
with $\pi$ as the conjugate field momentum and $m$ as the mass of the 
Klein-Gordon
particles.

The three parameters $a, \omega_z$ and $\omega_x$ characterize the
motion of a reference particle following the space-time curve $C$.
Thus, $a$ is the proper acceleration, while $\omega^z$ and $\omega^x$
specifies the angular velocity of the coordinate frame of the
trajectory, as viewed from the non-accelerated Minkowski frame. In
geometrical terms the three parameters describe the curvature and
torsion of the space-time curve, as discussed by Letaw and Pfautsch.
In the following we will assume $C$ to be a stationary curve, and
that means that all three parameters are independent of the time
coordinate along $C$. Due to Lorentz invariance the accelerated
Hamiltonian is then a time-independent operator.

The different types of stationary curves $C$ can be characterized by the
parameters $a, \omega_z$ and $\omega_x$, and following Letaw we
distinguish between the following types of curves which correspond to
qualitatively different types of motion (for simplicity we assume all
parameters to be positive),

\begin{enumerate}
\item{$a = \omega_z = \omega_x = 0$}\\
This corresponds to non-accelerated motion, \ie a linear space-time curve,
and will not be discussed further.

\item{$\omega_z = \omega_x = 0, a \neq 0 $}\\
This describes a curve with linear uniform acceleration, \ie a hyperbolic
space-time curve, here restricted to the ($t,x$) plane.

\item{$\omega_z =0,  \omega_x > a \neq 0 $}\\
This is a curve in the three-dimensional ($t,x,y$) hyperplane and can be
viewed as circular motion in a properly chosen coordinate frame.

\item{$\omega_z =0, \omega_x = a \neq 0 $}\\ This is also a curve in
the three-dimensional ($t,x,y$) hyperplane, and can be viewed as the
limiting case of circular motion where the speed tends to the speed of
light. This case is a limiting case between $\omega_x > a$ and
$\omega_x < a$. As we will show, the transition between the two
regimes is smooth, so we will not treat the limiting case separately.

\item{$\omega_z =0,  \omega_x < a \neq 0 $}\\
This is a curve in the three-dimensional ($t,x,y$) hyperplane which cannot
be viewed as circular motion. In a properly chosen coordinate frame the
($x,y$) projection of the curve is a hyperbolic cosine.

\item{$\omega_z \ , \omega_x \neq 0,  a \neq 0 $}\\
This is a curve that is not restricted to a hyperplane in Minkowski
space. In a properly chosen coordinate frame it can be viewed as circular
motion in the ($y,z$)-plane superimposed on a linear acceleration in the
x-direction.
\end{enumerate}

\vskip5mm

In the above we have focused on the properties of the reference curve
$C$.  This curve we may associate with the orbit of an accelerated
physical system or an accelerated "observer". More specifically we 
may view it as
the trajectory of a particle detector that probes the (Minkowski) 
vacuum state in
the accelerated frame. Although the reference trajectory has no 
extension in the
space-like directions the accelerated detector does not need to be 
pointlike, but
only small in order to avoid complications due to the singularities of the
accelerated coordinate system.

One should note that any curve defined by fixed space coordinates
in the accelerated coordinate system is a stationary curve, not only the
reference curve $C$. However, only for $C$ does the coordinate time coincide
with the proper time of the space-time curve, while for other curves there
is a constant scale factor relating the proper time to the
coordinate time. For other stationary curves than $C$, one should 
also note that
the time coordinate axis is no longer orthogonal to the space axes, as a
consequence of the rotation. In fact for a reference curve with rotation the
stationary curves at sufficiently large distance from $C$ may change from
time-like to space-like due to the rotation.

The collection of all the stationary curves of the accelerated
coordinate system can be viewed as defining a vector field, a Killing
vector field associated with a one-parameter symmetry of the metric. The
Hamiltonian \pref{accham2}, although primarily defined as
the time evolution operator of the quantum system, can also be
interpreted as the generator of this one-parameter set of space-time
transformations. Thus an explicit expression for the vector field is
given by the correspondences
\begin{eqnarray}
-i{H} & \longleftrightarrow & \vec{e}_{t} = \partdiff{}{t} \nn
i{K}_i & \longleftrightarrow & x^i \vec{e}_{t} + t \vec{e}_{i} =
x^i \partdiff{}{t} + t \partdiff{}{x^i} \nn
i{J}_i &
\longleftrightarrow & \epsilon_{ijk}\,  x^j \vec{e}_{k}  = 
\epsilon_{ijk}\,  x^j \partdiff{}{x^k} 
\end{eqnarray}
These expressions are useful for the further discussion of the causal
structure of space-time as seen in the accelerated frames. In the following
sections we shall also examine the definitions of vacuum states and particle
excitations in these frames.  We will in particular focus on the types of
motion where an event horizon exists. These are the cases 2), 5) and 6) in the
list.

\section{Linear acceleration and the Unruh effect}
\label{LinAcc}
\subsection{Rindler coordinates}
\label{Rincoord}
We will first consider the case of uniform acceleration $a$ with
no rotation, $\omega_x=\omega_z=0$. Although this is a well-studied case,
it gives the necessary background for the discussion of the other types of
motion where rotation is involved.

We choose in this case, as in all other cases, the Cartesian coordinate
frame
$(x^0 ; x^1, x^2, x^3) = (t; x, y, z)$ of Minkowski space to coincide
with the \textit{instantaneous (inertial) rest frame}
of the observer at proper time $\tau = 0$. The acceleration $a$
points along the
$x$-axis.
In this case the vector field corresponding to the stationary curves
of the accelerated coordinate system (\ie the integral curves of
the Killing vector field) is given by the correspondence
\begin{equation}
-i{H}_{O}\; \longrightarrow \; \vec{e}_{\tau} \, = \, \left ( 1 + ax
\right ) \vec{e}_{t} \, + \, at \, \vec{e}_{x}
\end{equation}
We note that by a shift of origin of the $x$-coordinate, $x' = x + 1/a$,
which gives
\begin{equation}
\vec{e}_{\tau} = a \left ( x' \vec{e}_{t} + t \vec{e}_{x'} \right )
\end{equation}
a simplification of the Hamiltonian is obtained,
\be{Ham2}
{H}_{O} = -a {K}_{x'}
\ee
with ${H}$ absorbed into the boost generator.

The accelerated coordinate system, which coincides with the Cartesian
coordinate system at $t=0$, is given by the
hyperbolic cylindrical coordinates
$(\tau,\xi, y, z)$
  with $\tau$ and $\xi$ defined by:
\be{Rindler}
t = \xi \sinh a\tau \; ,\quad  x = \xi \cosh a\tau
\ee
These coordinates are often referred to as the {\em Rindler
coordinates} \cite{Rindler66}. The reference curve
$C$ corresponds to
$\xi=1/a$, and has proper time equal to the coordinate time $\tau$. For
the case of linear acceleration, as for all other cases to be discussed,
the accelerated coordinate system is defined so that the hyperplanes
characterized by a constant coordinate time
$\tau$ correspond to planes of simultaneity for an observer accelerated
along
$C$.

\begin{figure}
\includegraphics[width=8cm,height=7cm]{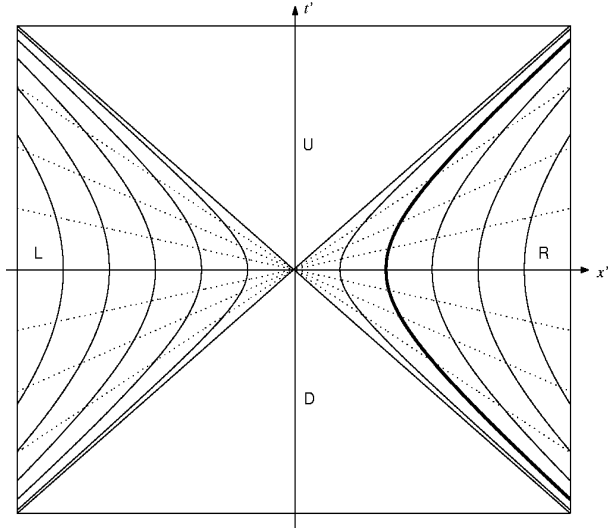}
\caption{The Rindler coordinate system. Dotted lines are integral
lines of $\vec{e}_{\xi}$ and solid lines are integral curves of
$\vec{e}_{\tau}$. The thick solid line is the stationary reference 
curve $C$ or the
world line of the uniformly accelerated ``observer".}
\label{AcceleratedSystem}
\end{figure}

As is well known, the accelerated Rindler coordinate system associated
with uniform linear acceleration covers only a part of the full Minkowski
space. As shown in Fig.\ref{AcceleratedSystem} the part that is 
covered includes the
``right Rindler wedge" R, where $\tau$ increases in the positive time 
direction,
and the ``left Rindler wedge" L, where $\tau$ runs backwards in time.
The presence of ``backward running" stationary curves in the Rindler
coordinate system is related to the presence of negative energy
excitations in the accelerated reference frame.

Viewed from an observer accelerated along the curve $C$ (or any other
curve with constant $\xi$) in R, the region L is causally disconnected,
since no light signal emitted from $C$ can reach L and no light signal
from L can reach $C$. The limiting point (hypersurface) $\xi =0$
between R and L defines an event horizon viewed from $C$. The regions D
and U, which are excluded from the Rindler coordinate system are
not causally disconnected, but viewed from $C$ the region D lies in the
infinite past and U in the infinite future. Thus, D may be viewed as being
separated from R by a past horizon and U by a future horizon. This
causal structure of the Rindler coordinate system is the underlying
reason that a quantum field theory restricted to the right Rindler wedge
can be defined.

The boost operator, viewed as the Hamiltonian \pref{Ham2} of the accelerated
frame, is different from the original Hamiltonian in one important aspect: It
has a spectrum that is not bounded from below. (That is true not only for the
special case of linear acceleration, but for the general case of accelerated
motion described by the Hamiltonian \pref{accham}.) The spectrum is in fact
symmetric under change of sign as follows from the anti-commutation relation
with space inversion,
\be{Pcom}
P K_x=-K_x P
\ee
As a result a {\em vacuum state} defined as the
ground state of the Hamiltonian \pref{Ham2} does not exist.

Even without a well-defined ground state the free field
Hamiltonian can be written in the conventional form
\be{convham}
H_O=\sum\limits_i\epsilon_i\; a^{\dag}_i a_i
\ee
with $a^{\dag}_i$ and $a_i$ as creation and annihilation operators, \ie
which satisfy the standard commutation relations
\be{standcom}
\com{a_i}{a^{\dag}_j}=\delta_{ij}\;,\;\;\com{a_i}{a_j}=\com{a^{\dag}_i}{a^{\dag}_j}=0
\ee
The operators $a^{\dag}_i$ can be defined as a linear combination
of the usual creation operators associated with free particles in
Minkowski space (and $a_i$ as the corresponding linear combination of
annihilation operators). This follows from the fact that the (free field)
expressions for the generators of Lorentz transformations are bilinear in
creation and annihilation operators and can be diagonalized without
mixing these two types of operators.

With the Hamiltonian written in the form \pref{convham} a (generalized)
vacuum state can be defined as the state annihilated by the operators
$a_i$,
\be{vac}
a_i\ket{0}=0
\ee
and the full space of states can be generated in the standard way by repeated
application of the creation operators on the vacuum state. The operators
$a_i^{\dag}$ create excitations (particles) with energy
$\epsilon_i$ with reference to the accelerated coordinate frame, and since
$\ket{0}$ is not the ground state of the Hamiltonian, excitations both
with positive and negative energy $\e_i$ can be created from the vacuum
state
$\ket{0}$. Note that when the operators $a_i$ are defined as linear
combinations of the standard (Minkowski space) annihilation operators, the
generalized vacuum of the accelerated frame is identical to the ordinary
Minkowski vacuum.

\subsection{The transformed vacuum}
\label{transvac}
In the above description of the field theory vacuum no particles are present in
the physical (Minkowski) space. The Unruh effect in this picture is understood
as due to processes where the accelerated detector is excited by emitting
negative energy particles. However, the symmetry of the spectrum of
$H_O$ between positive and negative values allows alternative definitions of
the vacuum state and the particle excitations. A particular choice gives rise
to the Fulling-Rindler vacuum, where all excitations restricted to the right
Rindler wedge $R$ have positive energy. In this picture the Minkowski vacuum
state contains  particles relative to the accelerated vacuum
state and the Unruh effect is described as due to
absorption of such (positive energy) particles.

To discuss this point we focus on
the field operator of the (real) Klein-Gordon field. When expressed in terms
of the Rindler coordinates, it can be written as
\be{fieldop}
\phi(\xi,y,z,\tau)=\suml_i(e^{-i\e_i\tau} f_i(\xi,y,z) a_i +
e^{i\e_i\tau} f_i^*(\xi,y,z) a^{\dag}_i\,)
\ee
with $f_i(\xi,y,z)\; a_i$ and $f_i^*(\xi,y,z)$ as stationary solutions (with
respect to $\tau$) of the Klein-Gordon equation,
\be{KGeq}
\big(\frac{1}{\xi}\,\pa{\xi}\,\xi\,\pa{\xi}+\pa{y}^2+\pa{z}^2+(\e_i^2-m^2)\big)
f_i(\xi,y,z)=0
\ee
To simplify the notation we have labelled the solutions $f_i$
with a single (discrete) index $i$, which in reality should be 
replaced by a set of
(continuous) momentum-energy variables that specify the solution. The 
solutions can
be separated into those with positive norm and those with negative 
norm with respect
to the scalar product of the (classical) Klein-Gordon field, which in Rindler
coordinates has the form
\be{scalprod}
\langle f , h \rangle = i \int \frac{1}{a\xi}\left ( f^*(\xi,y,z)\;
\pa{\tau}h(\xi,y,z) - \pa{\tau}f^*(\xi,y,z)\; h(\xi,y,z) \right )
\mathrm{d}\xi\mathrm{d}y\mathrm{d}z
\ee
Consistency between the requirement of standard commutation relations
for $a_i$ and $a_j^{\dag}$ and canonical commutation rules for $\phi$ and
its conjugate momentum implies that in the expansion \pref{fieldop} the
functions $f_i$ are positive norm solutions and correspondingly $f_i^*$
are negative norm solutions.
Note that positive norm no
longer corresponds to positive frequency, as is the case for
Minkowski space quantization.

In the sum included in Eq. \pref{fieldop} the energies $\e_i$ take
both positive and negative values. Due to the symmetry of the the
spectrum it can, however, be rewritten as a sum only over positive
energies in the following way,
\be{fieldop2}
\phi(\xi,y,z,\tau)&=& \suml_{i\,(\e_i>0)}[(e^{-i\e_i\tau} f_i(\xi,y,z) a_i +
e^{+i\e_i\tau} g_i(\xi,y,z) b_i)\nn &&\hskip1cm+
(e^{i\e_i\tau} f_i^*(\xi,y,z) a^{\dag}_i+e^{-i\e_i\tau} 
g_i^*(\xi,y,z) b^{\dag}_i\,)]\nn
&=& \suml_{i\,(\e_i>0)}[e^{-i\e_i\tau}( f_i(\xi,y,z) a_i+ g_i^*(\xi,y,z)
b^{\dag}_i )\nn &&\hskip1cm+
e^{i\e_i\tau} (f_i^*(\xi,y,z) a^{\dag}_i+ g_i(\xi,y,z) b_i\,)]
\ee
In this expression we have explicitly separated out the negative 
energy solutions, which we refer
to as $g_i$, and introduced the notation  $b_i$ and $b_i^{\dag}$ for 
the corresponding annihilation
and creation operators.

Since the operators $a_i$ and $b_i^{\dag}$ have the same 
time-dependent prefactor, a
Bogoliubov transformation may be performed,
\be{Bog}
a_i&=&\tilde a_i \cosh\chi_i + \tilde b_i^{\dag} \sinh\chi_i\nn
b_i^{\dag}&=&\tilde a_i  \sinh\chi_i + \tilde b_i^{\dag} \cosh\chi_i
\ee
The transformed operators $\tilde a_i$ and $\tilde b_i^{\dag}$ and 
their hermitian
conjugates satisfy the same commutation relations as the original creation and
annihilation operators and the field operator expressed in terms of 
the new ones has
the same form as before,
\be{fieldop3}
\phi(\xi,y,z,\tau)&=& \suml_{i\,(\e_i>0)}[e^{-i\e_i\tau}( \tilde f_i(\xi,y,z)
\tilde a_i+ \tilde g_i^*(\xi,y,z) \tilde b^{\dag}_i )\nn &&\hskip1cm+
e^{i\e_i\tau} (\tilde f_i^*(\xi,y,z) \tilde a^{\dag}_i+ \tilde g_i(\xi,y,z)
\tilde b_i\,)]
\ee
with transformed fields
\be{transfield}
\tilde f_i(\xi,y,z)&=&f_i(\xi,y,z) \cosh\chi_i +  g_i^*(\xi,y,z)
\sinh\chi_i\nn
\tilde g_i(\xi,y,z)&=& f_i^*(\xi,y,z)\sinh\chi_i+ g_i(\xi,y,z) \cosh\chi_i
\ee

A new generalized vacuum state may be defined relative to the 
transformed operators,
\be{newvac}
\tilde a_i\ket{\tilde 0}=0\,,\,\;\tilde b_i\ket{\tilde 0}=0
\ee
and this is different from the Minkowski vacuum due to the mixing of 
creation and
annihilation operators. In fact a continuum of different vacuum states 
can be defined in
this way, but like the Minkowski vacuum they generally suffer from 
the defect that both
positive and negative energy excitations (of the accelerated 
Hamiltonian) may be created
from the state $\ket{\tilde 0}$. The Minkowski vacuum will now 
contain field quanta
relative to the transformed vacuum and for a detector that is
stationary in the accelerated frame there will therefore be two sources 
for excitations.
The detector can either be excited by absorbing positive energy 
quanta or by emitting
negative energy quanta.

A particular transformation can be chosen where all negative energy 
field quanta are
located beyond the event horizon, \ie in the left Rindler wedge $L$ ($\xi<0$).
This gives rise to the Rindler or Fulling vacuum, which is a true 
vacuum state in the
sense that it is the ground state of $H_O$, provided all states are 
restricted to the
right Rindler wedge
$R$. This possibility can be viewed as due to the symmetry of the 
theory under $PCT$,
which gives rise to a relation between the functions $f_i$ and $g_i$ on the two
sides of the horizon \cite{Sewell82,Hughes85,Bell85}. We will give a simple
demonstration of this.

The $PCT$ transformation $\theta$ acts on the field operator in the 
following way,
\be{PCT}
\theta \phi\,(\xi,y,z,\tau) \,\theta^{-1}= \phi^{\dag}(-\xi,-y,-z,\tau)
\ee
Like the parity operator, $\theta$ anticommutes with the boost
operator and therefore maps positive energy solutions into negative energy
solutions,
\be{PCT2}
\theta a_i \,\theta^{-1}=b_i\;,\;\;\theta b_i \,\theta^{-1}=a_i
\ee
Since $\theta$ acts as an antilinear operator this gives the following relation
between the functions $f_i$ and $g_i$,
\be{fgrel}
g_i(\xi,y,z)=f_i^*(-\xi,-y,-z)
\ee

We further note that the definition of the Rindler coordinates implies that the
strong space-time reflection $PT$ can be performed {\em either} by an 
inversion $\xi
\ra-\xi$ (together with $y \ra -y, z \ra -z$) {\em or} by an analytic 
continuation
$\tau
\ra \tau \pm \pi i/a$.  This is related to the fact that $PT$ can be 
viewed as a
complex extension of the Lorentz transformations (which here induce 
translations in
$\tau$) \cite{Bell55}.
As a result $f_i(-\xi,-y,-z)$ can be related to $f_i(\xi,y,z)$ by an analytic
continuation of the full time-dependent solution of the Klein-Gordon equation.

Since the operators $a_i$ and $b_i$ have been introduced without mixing the
original Minkowski space annihilation operators with creation 
operators, this means
that the corresponding functions $f_i$ and $g_i$ include only positive
frequency components with respect to Minkowski time. We can therefore
expand $f_i$ as follows,
\be{fexpand}
e^{-i\e_i\tau}f_i(\xi,y,z)&=&\int d^3 p F_i(\vec p) e^{-i(\omega_p t - 
\vec p\cdot \vec
x)}\;, \;\;\; \omega_p=\sqrt{p^2+m^2}\nn
&=&\int d^3 p F_i(\vec p) e^{i(p_y y+p_z z)}e^{-i\xi(\omega_p \sinh 
a\tau - p_x\cosh a\tau)}
\ee
A similar expression is valid for $g_i$.  For a complex extension
$\tau\ra\tau+i\vartheta/a$, the change of the exponential, factor is
\be{expfac}
e^{-i\xi(\omega_p \sinh a\tau - p_x\cosh a\tau)}\ra 
e^{-i\xi\cos\vartheta(\omega_p \sinh
a\tau - p_x\cosh a\tau)}e^{\xi\sin\vartheta(\omega_p \cosh
a\tau - p_x\sinh a\tau)}
\ee
where the last factor determines whether the integral \pref{fexpand} 
is convergent or
divergent for large $p$. Since $\omega_p>p$, convergence implies that 
for positive
$\xi$ the Rindler time $\tau$ should be analytically continued in the 
lower complex
half plane, while for negative $\xi$ it should be continued in the 
upper half plane.
With this prescription for the analytic continuation of the solutions to the
space-time inverted point we find
\be{PTinvert}
f_i(-\xi,-y,-z)&=&e^{-\frac{\e_i\pi}{a}}f_i(\xi,y,z)\; , \;\; \xi>0\nn
g_i(-\xi,-y,-z)&=&e^{\frac{\e_i\pi}{a}}\;g_i(\xi,y,z)\; , \quad \xi>0
\ee
  which, together with \pref{fgrel}, finally leads to the following 
relations between $f_i$ and
$g_i$ in the same Rindler wedge
\be{fgrel2}
f_i(\xi,y,z)=e^{+\frac{\e_i\pi}{a}}\;g_i^*(\xi,y,z)\;,\quad\xi>0\nn
f_i(\xi,y,z)=e^{-\frac{\e_i\pi}{a}}\;g_i^*(\xi,y,z)\;,\quad\xi<0
\ee
Thus, if we define transformed fields as
\be{transfield2}
\tilde f_i(\xi,y,z)&=&\sinh(\frac{\e_i\,\pi}{2a})^{-\half}
[e^{+\frac{\e_i\pi}{2a}}f_i(\xi,y,z)-e^{-\frac{\e_i\pi}{2a}}\;g_i^*(\xi,y,z)]\nn
\tilde g_i(\xi,y,z)&=&\sinh(\frac{\e_i\,\pi}{2a})^{-\half}
[e^{+\frac{\e_i\pi}{2a}}g_i(\xi,y,z)-e^{-\frac{\e_i\pi}{2a}}\;f_i^*(\xi,y,z)]
\ee
which corresponds to a transformation of the form \pref{transfield} with
\be{chsh}
\cosh\chi_i=\frac{e^{+\frac{\e_i\pi}{2a}}}{\sqrt{2\sinh(\frac{\e_i\,\pi}{a})}}\;,\;\;
\sinh\chi_i=-\frac{e^{-\frac{\e_i\pi}{2a}}}{\sqrt{2\sinh(\frac{\e_i\,\pi}{a})}}
\ee
then the transformed field
$\tilde f_i$ vanishes identically for
$\xi<0$ and
$\tilde g_i$ vanishes identically for $\xi>0$. Thus, for this value for the
transformation parameter $\chi_i$ the field modes of the two sides of the
horizon are completely decoupled.

In terms of the new creation and annihilation operators Minkowski vacuum
contains (particle) excitations,
\be{vacexcit}
\mean{\tilde a_i^{\dag} \tilde a_i}_M&=&\sinh^2\chi_i\mean{b_i b_i^{\dag}}_M\nn
&=& \frac{e^{-\frac{\e_i\pi}{a}}}{2\sinh(\frac{\e_i\,\pi}{a})}\nn
&=&\frac{1}{e^{\frac{2\pi\e_i}{ a}}-1}
\ee
The expression corresponds to a thermal Bose-Einstein distribution over the
excited levels, with temperature given by
\be{TU}
k_B T_a= \frac{ a}{2\pi}
\ee
which is the {\em Unruh temperature}.

\section{Planar motion with acceleration and rotation}
We now consider motion along stationary
worldlines where the acceleration is no longer linear. In these cases the
co-moving, Fermi-Walker transported frame will be rotating relative to an
{\em inertial} rest frame. Thus, $\bs\omega\neq 0$, and the general
Hamiltonian is given by
\pref{accham2}. We restrict, in this section, the motion to be planar, with
the Hamiltonian given by
\be{acchamplane}
{H}_{O} \, = \, {H} - a{K}_x - \omega {J}_z
\ee
It is convenient to simplify the Hamiltonian by making a boost in the
$y$-direction in addition to a shift of origin in the $x$-direction. With
$\beta$ as the velocity parameter the transformation from the lab frame to
the boosted coordinates is
\be{boost}
y'=\gamma(y-\beta t) \;,\;\;\;x'=x+\frac{a}{a^2-\omega
^2}\;,\;\;\;z'=z\;,\;\;\; t'=\gamma(t-\beta y)
\ee
and the Hamiltonian expressed in terms of operators of the transformed
frame is
\be{transh}
{H}_{O} = \gamma \left [ \,-\frac{\omega(\omega-a\beta)}{a^2-\omega^2}
{H}' +
\frac{\omega(a-\omega\beta)}{a^2-\omega^2}
{P}_{y'} -
\left ( a - \beta\omega \right ) {K}_{x'} - \left
( \omega - \beta a \right ) {J}_{z'} \,
\right ]
\ee
  with $P_{y'}$ as the translation operator in the $y'$-direction.

We note that
when $a>\omega$ the velocity parameter can be chosen as  $\beta=\omega/a$ so
that the coefficients of $H_O$ and $J_z$ vanish. The Hamiltonian then 
simplifies to
\be{transh2}
{H}_{O} = -\sqrt{a^2 - \omega^2}\,{K}_{x'} + \frac{\omega}{\sqrt{a^2 -
\omega^2}}
\,{P}_{y'} \;,\;\;\;\;a>\omega
\ee If instead $a<\omega$ it can be
chosen as $\beta=a/\omega$ so that the coefficients of $K_{x'}$ and $P_{y'}$
vanish. The form of
the Hamiltonian then is
\be{transh3}
{H}_{O} = \frac{\omega}{\sqrt{\omega^2 -
a^2}}
\,{H}' - \sqrt{\omega^2 - a^2}\,{J}_{z'} \;,\;\;\;\;a>\omega
\ee
We shall discuss these two cases separately. For simplicity we suppress in this
section the $z$-coordinate, since the motion is restricted to the 
$(x,y)$-plane.

\subsection{Accelerated coordinates for planar trajectories with rotation}
{\em The case $ a>\omega$\;\; }

The stationary curves generated by the time evolution operator 
\pref{transh2} are
described by the (tangent) vector field
\be{vecfield1}
\vec{e}_{\tau} = \sqrt{a^2 - \omega^2}
\left ( x' \vec{e}_{t'} + t' \vec{e}_{x'} \right )-
\frac{\omega}{\sqrt{a^2 - \omega^2}} \vec{e}_{y'}
\ee
which is here expressed in terms of the coordinates of the boosted 
inertial frame. The
expression shows that the trajectory projected on the $(t',x')$-plane 
is the same as
for hyperbolic motion, \ie for linear motion with constant proper 
acceleration. In
addition there is motion in the $y'$ direction, and measured with the 
time coordinate
$\tau$ of the accelerated frame, this component of the motion 
corresponds to a drift
with constant velocity.  However, measured with the inertial time $t'$ the
velocity in the $y'$-direction will slow down with increasing $|t'|$ 
due to time
dilatation. This means that asymptotically (for $t'\ra\pm
\infty$) the motion will be dominated by the $x'$-component.

It is interesting
to note that there is no rotation involved in the motion as seen from the
boosted inertial frame. The rotation of the curve, as measured in the
local inertial rest frame, can be seen as due to
composition of boosts in two different directions, in the $x'$-direction and
the $y'$-direction, when transforming to the rest frame of the
stationary curve.

The decomposition of the motion in the $x'$ and $y'$ directions
motivates the introduction of a new set of Rindler coordinates which are
adjusted to the component with linear acceleration in the $x'$ direction,
\be{Rindler2}
t' &= \xi' \sinh \sqrt{a^2 - \omega^2}\tau'\;,\;\;
    x' = \xi' \cosh \sqrt{a^2-\omega^2}\tau'
\ee
In terms of these Rindler coordinates the vector field \pref{vecfield1}
takes the form
  \be{vecfield2}
\vec{e}_{\tau} = \sqrt{a^2 - \omega^2}
\vec{e}_{\tau'}-
\frac{\omega}{\sqrt{a^2 - \omega^2}} \vec{e}_{y'}
\ee

Since we now in reality are working with four different coordinate systems,
we make the following reminder: The original inertial frame with coordinates
$(t,x,y)$ coincides with the inertial rest frame of the accelerated
world line at time $t=0$. If we denote by $(\tau,\xi,\eta)$ the accelerated
coordinates, we may therefore identify $\xi$ with $x$ and $\eta$ with $y$
at $t=\tau=0$. The boosted inertial frame, with coordinates $(t',x',y')$
moves with constant velocity along the $y$-axis. This frame is useful
since it decomposes the motion in two independent components in the $x'$-
and $y'$-directions. Finally the Rindler coordinate system, with
coordinates $(\tau',\xi', y')$ is the rest frame of the $x$-component of
the motion. It is not identical to the accelerated frame of the
stationary orbit, since it does not take into account the motion in the
y-direction. We note that the $z$ coordinate is common for all coordinate
systems since all motion takes part in the $x,y$ directions only.

The reference orbit $C$ is defined as the stationary curve where the coordinate
time $\tau$ coincides with the proper time, \ie $\vec e_{\tau}\cdot\vec
e_{\tau}=-1$. From the expression \pref{Rindler2}, we find the Rindler
coordinates of this orbit to be, when expressed with the proper time $\tau$
as curve parameter
\begin{align}
\label{orbit}
\tau'(\tau) &=  \, \tau \nn
\xi'(\tau) &= \frac{a}{a^2 - \omega^2} \notag \nn
y'(\tau) &= - \frac{\omega}{\sqrt{a^2 - \omega^2}} \, \tau \nn
z(\tau) &= 0
\end{align}
The $\xi'$ coordinate is constant, but there is a constant drift in the $y'$
direction.

The causal structure of space time as seen from the accelerated orbit is
most easily discussed in the boosted inertial frame. Since the asymptotic
behaviour of the stationary curve is dominated by the $x$-component of
the motion, there are event horizons located in the same positions as for
linear accelerated motion. There is a past horizon at $x'=-ct'$, so that no
light signal emitted from the accelerated orbit can reach points
with $x'<-ct'$, and there is a future horizon at $x'=ct'$ so that no
light signal emitted from a point $x'<-ct'$ can reach the accelerated
orbit. The location of these horizons in the rest frame of the
accelerated orbit can be found by use of the identification of space
coordinates of this frame with those of the (original) inertial frame at $t =
0$. The Lorentz transformation \pref{boost} gives the locations at
\be{horizons}
\eta=\pm \frac{\sqrt{a^2-\omega^2}}{\omega}(\xi+\frac{a}{a^2-\omega^2})
\ee
where ``+" gives the past horizon and ``-" gives the future
horizon. Note that this expression is valid
not only for $\tau=0$, but for all $\tau$, since the situation in the
accelerated frame is {\em stationary}. In the same way as for the horizon of a
linearly accelerated orbit the horizons here correspond to singularities of the
accelerated coordinate system, where the time coordinate is ill-defined.

\begin{figure}
\includegraphics[width=8cm,height=7cm]{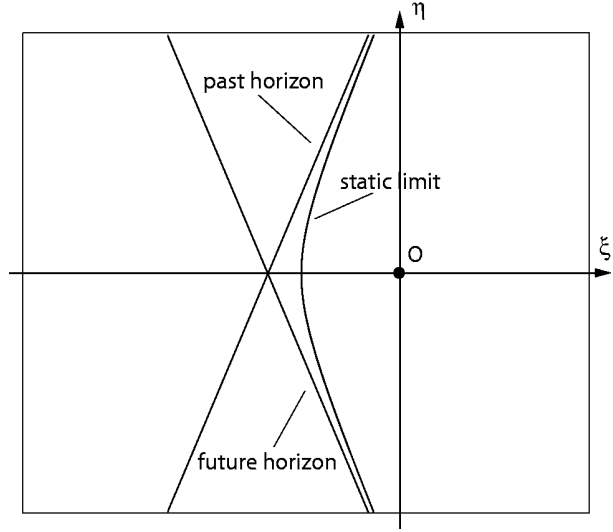}
\caption{Event horizons and static limit in the accelerated frame. The
stationary reference curve $C$ corresponds to the origin $O$. The two
directions in the plane correspond to the two space-like directions where
$\xi$ coincides with $x$ of the fixed inertial frame and $\eta$ coincides
with $y$ for $t=0$. The distances to the horizon and to the
static limit are determined by the two parameters $a$ and $\omega$.}
\label{fig:Horizon}
\end{figure}

Since the accelerated frame is rotating, there exists a {\em static limit} in
this coordinate system. This is the limit for where physical objects can be
{\em stationary} with respect to the coordinate system. Beyond
the static limit the stationary curves become space-like rather than time-like. The location of the static limit can be found by  considering the
norm of the coordinate vector $\vec e_{\tau}$ as a function of the
coordinates,
\be{statlim}
\vec{e}_{\tau} \cdot \vec{e}_{\tau} =
\vec{e}_{\tau'} \cdot \vec{e}_{\tau'} + \frac{\omega^2}{a^2 - \omega^2}
\, \vec{e}_{y'} \cdot \vec{e}_{y'} = \left ( a^2 - \omega^2 \right )
({x'}^2-{t'}^2) - \frac{\omega^2}{a^2 - \omega^2}
\ee
The norm changes from negative to positive at the static
limit. In order to find its location in terms of the coordinates of
the accelerated frame, we again make use of the fact that the coordinates
$(\xi,\eta)$ coincide with $(x,y)$ for $t=0$ (or equivalently for $\tau
=0$). By use of the Lorentz transformation \pref{boost} we find for $t=0$
(and with $(x,y)$ replaced by $(\xi,\eta)$),
\be{statlim2}
\vec{e}_{\tau} \cdot \vec{e}_{\tau} = \left ( a^2 - \omega^2 \right )
\Big[(\xi+\frac{a}{a^2-\omega^2})^2-\eta^2\frac{\omega^2}{a^2-\omega^2}\Big]
  - \frac{\omega^2}{a^2 -\omega^2}
  \ee
The scalar product vanishes for
\be{statlim3}
  \left ( \frac{a^2 - \omega^2}{\omega} \right )^2
(\xi+\frac{a}{a^2-\omega^2})^2-(a^2-\omega^2)\eta^2
=1
\ee
This defines a hyperbola in the $(\xi,\eta)$-plane, with the event horizons
as asymptotes. Viewed from the stationary orbit $C$, which is
given by $(\xi,\eta)=(0,0)$, the event horizons are located {\em behind}
the static limit.

We note that when $\omega\ra 0$, where the case of the linear
acceleration is recovered, both the event horizons tend to the same
straight line $\xi\ra-1/a$ and the static limit merges with the horizons.

When the rotation instead increases and $\omega\ra a$, the distance to both
horizons go to infinity and they therefore disappear from the
$(\xi,\eta)$-plane. The static limit on the other hand stays at a finite
distance and changes to a parabola. \\

\noi
 {\em The case $ a<\omega$}

In this case the (accelerated) time translation operator is given by
\pref{transh3} and the corresponding vector field is
\be{vecfield3}
\vec{e}_{\tau} = \frac{\omega}{\sqrt{\omega^2 -
a^2}}
\,\vec{e}_{t'} + \sqrt{\omega^2 - a^2}\,(x'\vec{e}_{y'}-y'\vec{e}_{x'})
\ee
This describes {\em circular motion}, as can be seen explicitly by changing to
cylindrical coordinates
\be{cylcoord}
x' = r \cos \theta \;\;\; y' = r \sin \theta
\ee
which gives
\begin{equation}
\vec{e}_{\tau} \, = \, \frac{\omega}{\sqrt{\omega^2 - a^2}}
\vec{e}_{t'} + \sqrt{\omega^2 - a^2}\; \vec{e}_{\theta}
\end{equation}
At $\tau = 0$, the reference curve $C$ is located at $x' = -a/(\omega^2 -
a^2)$, corresponding to $r = a/(\omega^2 - a^2)$ and $\theta = \pi$. 
Expressed as a
function of the proper time coordinate $\tau$, the trajectory is
\begin{align}
t' &= \frac{\omega}{\sqrt{\omega^2 - a^2}} \, \tau \nn
r &= \frac{a}{\omega^2 - a^2}  \nn
\theta &= \pi + \sqrt{\omega^2 - a^2} \, \tau
\end{align}
This space-time curve describes motion in a circle of radius $r 
=a/(\omega^2 - a^2)$
with angular velocity $\Omega=(\omega^2-a^2)/\omega$ relative to the 
boosted inertial
frame.

In this case there clearly is no event horizon since the orbit is 
restricted to a
bounded region of space, so that a signal from an event anywhere in 
space will be able
to reach the space-time curve and any space-time point can be reached 
by a light signal
from a point on the curve. However, also for circular motion there is 
a static limit defined by the accelerated coordinate system. The norm of the
vector
$\vec{e}_{\tau}$ now is
\begin{equation}
\vec{e}_{\tau} \cdot \vec{e}_{\tau} \, = \, \frac{\omega^2}{\omega^2 -
a^2} - r^2 \left ( \omega^2 - a^2 \right )
\end{equation}
hence, $\vec{e}_{\tau}$ becomes spacelike for $r >
\omega/(\omega^2 - a^2)$, which is the radius where the velocity of 
the rotating
coordinate system exceeds the velocity of light. The location of this 
static limit in
the accelerated coordinates can be found in the same way as before, 
that is by
performing a boost to the inertial frame that coincides with the 
rotating frame at $t=0$
and by making the identification $(x,y)\ra(\xi,\eta)$ for $t=0$. The 
location is
determined by ,
\be{statlimrot}
\left ( \frac{\omega^2 - a^2}{\omega} \right )^2 \left ( \xi -
\frac{a}{\omega^2 - a^2} \right )^2 +
\left ( \omega^2 - a^2 \right ) \eta^2 \, = \, 1
\ee
which is the equation for an ellipse, with the major semiaxis in 
the $\xi$-direction
and the minor semiaxis in the $\eta$-direction. We can interpret this 
shape of the static limit, as a deformation of a circle of radius $\omega/(\omega^2-a^2)$ in the $(x',y')$-plane. 
The deformation is due to the length contraction in the $y'$ (or $\eta$)-direction introduced by the
transformation to the rest frame of the accelerated trajectory.

 From the above discussion it follows that there is a qualitative 
difference between
the cases $a>\omega$ and $a<\omega$. In the first case the (3-)space 
projection of
the accelerated trajectory is unbounded in any inertial frame, and 
the asymptotic
behavior implies that event horizons exist. In the latter case, the 
motion is bounded
and can be viewed as circular motion in a properly chosen inertial 
frame. No event
horizons exist in this case. The limiting case $a=\omega$ can be 
reached from both
sides. Viewed as a limiting case of circular motion it corresponds to the
ultra-relativistic limit where the velocity of circulation tends to 
the velocity of
light.

\begin{figure}
\includegraphics[width=7cm,height=7cm]{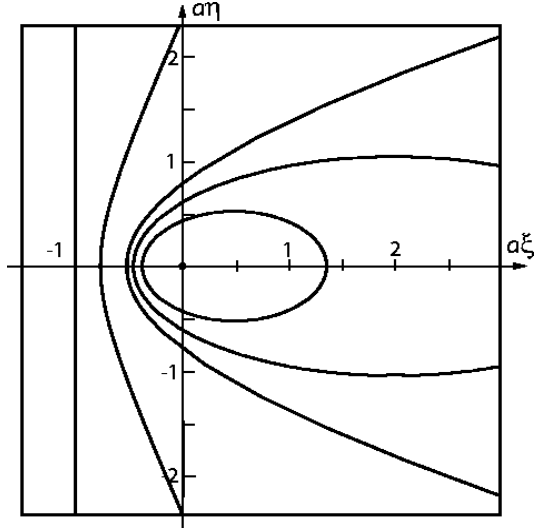}
\caption{The static limit for different values of $\omega/a$ as seen 
in the accelerated
coordinate frame. There is a continuous change from $\omega<a$ to 
$\omega>a$. From left
to right the curves in the
$(\xi,\eta)$-plane correspond to $\omega=0$ (straight line), 
$\omega=0.3 a$ (hyperbola),
$\omega=a$ (parabola), $\omega=1.2a$ and $\omega=1.4a$ (both 
ellipses). The coordinates
are measured in units of $1/a$. }
\label{static}
\end{figure}
One should note, however, that this qualitative difference has mainly 
to do with the
asymptotic form of the space time curves as viewed from an inertial 
frame. For a finite
time interval, or viewed in the co-moving frame, there is no singular 
behaviour at the
point
$a=\omega$. This is demonstrated in Fig.\ref{static} where the static limit as
viewed in the accelerated coordinate system, is shown for several 
different values of
$\omega/a$. The static limit changes continuously from a hyperbola 
(with the event
horizons as asymptotes) for
$a>\omega$ through a parabola for $a=\omega$ to an ellipse for $a<\omega$.

\subsection{Vacuum states in the accelerated frame}
We first consider the case $a>\omega$ with the Hamiltonian given by
\pref{transh2}. We note that since ${K}_{x'}$ and ${P}_{y'}$ commute,
the eigenstates of the Hamiltonian $H_O$ can also be chosen as eigenstates of the
boost operator ${K}_{x'}$ and the translation operator $P_{y'}$ . This means that
the Hamiltonian can be diagonalized in the same way as for linear acceleration.
Thus, also for $\omega\neq0$ and $a>\omega$ there exists a continuum of possible vacuum states
that are connected by Bogoliubov transformations which mix creation
and annihilation operators. We focus again on the particular transformation that
separate the field modes on the two sides of the horizons. The field
modes $\tilde f_i$ corresponding to negative eigenvalues of
$K_{x'}$ are the relevant ones for the reference curve $C$ in the
right Rindler wedge, since these are the ones that have positive norm
on that wedge. We denote these eigenvalues by $-\Omega$, with $\Omega>0$.

The field modes $\tilde f_i$ can be identified with the corresponding 
modes previously
discussed, except that they are now functions of the new
Rindler coordinates $(\tau',\xi',y',z)$. The explicit form is
\begin{equation}
\tilde f_{k_{y'}k_{z} \Omega}(\tau', \xi', y',z) =
\begin{cases}
\frac{\sqrt{\sinh(\pi\Omega)}}{2\pi^2}
e^{-i\Omega\sqrt{a^2-\omega^2}\tau'} e^{i(k_{y'} y'+k_{z} z)}  K_{i\Omega}(\Xi
\xi'), & \xi' > 0, \; \Omega > 0 \\
0, & \xi' < 0, \; \Omega > 0
\end{cases}
\label{f-function}
\end{equation}
where the index $i$ here is specified as the set of eigenvalues 
$\Omega$, $k_{y'}$ and
$k_z$ of the operators ${K}_{x'}$, ${P}_{y'}$ and ${P}_{z}$, and the
parameter
$\Xi$ is
\be{radial}
\Xi=\sqrt{k_y^2+k_z^2+m^2}
\ee
with $m$ as the mass of the Klein-Gordon particles. The function 
$K_{i\Omega}$, which is a
solution of the Klein-Gordon equation in Rindler coordinates \pref{KGeq}, is a
modified Bessel function (a MacDonald function of imaginary order)
\cite{MacDonald}.

Even if the wave functions are the same as for linear
acceleration one should note an important difference. For linear 
acceleration the
restriction of the wave functions to the accessible side of the event 
horizon will at
the same time be a restriction of the energies to positive values. 
Here this restriction means that
$\Omega>0$ for the relevant eigenvalue of the boost operator. 
However, this eigenvalue
is no longer proportional to the {\em energy} eigenvalue, since the 
energy also gets
contribution from
${P}_{y'}$. The eigenvalues of $H_O$ are
\be{energyrot}
\epsilon=\sqrt{a^2-\omega^2}\;\Omega+\frac{\omega}{\sqrt{a^2-\omega^2}}k_{y'}
\ee
For sufficiently large negative $k_{y'}$ the last term will dominate 
the first term
and give a negative energy eigenvalue.

\begin{figure}
\includegraphics[width=7cm,height=4cm]{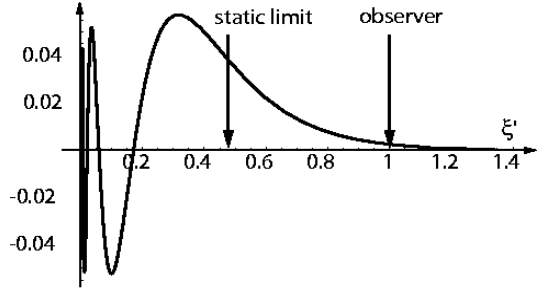}
\caption{Stationary solution of the Klein-Gordon equation in the Rindler
coordinates. The figure shows a solution with energy $\e=0$ for the 
Hamiltonian of
the accelerated coordinate system. For all negative values $\e<0$ the 
solutions have
only an exponential tail on the "physical" side of the static limit. }
\label{negenergi}
\end{figure}
It is of interest to note that these negative energy solutions will
essentially be located between the event horizon and the static 
limit. To see this
we first note that the static limit of the accelerated coordinate system, as
well as the horizons and the reference curve $C$ have fixed $\xi'$ coordinates,
even if the Rindler coordinate system does not define a rest frame for the
accelerated observer. Thus both horizons have Rindler coordinate $\xi_{ho}=0$,
the static limit is located at $\xi_{sl}=\omega/(a^2-\omega^2)$ and the
accelerated observer is located at $\xi'=a/(a^2-\omega^2)$. We next 
note that the energy
eigensolutions \pref{f-function} have an oscillatory behaviour for 
small $\xi'$ but are
exponentially damped for large $\xi'$. As shown by the Klein Gordon 
equation the
transition point between the two types of behaviour is 
$\xi'_{tr}=\Omega/\Xi$. For
negative energy states we have 
$\Omega<|k_{y'}|\omega/(a^2-\omega^2)$, as follows from
\pref{energyrot}, which implies the inequality
\be{ineq}
\xi'_{ho}<\xi'_{tr}<\xi'_{sl}
\ee
Thus, the negative energy solutions are are mainly located inside the 
static limit, but this
is not fully so due to the exponential tail of the wave functions, as 
illustrated in
Fig.\ref{negenergi}. The penetration of the negative energy states 
into the physical
region of the accelerated frame is important for excitations of an 
accelerated detector. The
modes can contribute to the excitations since negative energy quanta 
can be emitted into the
modes by the detector.

To summarize, for accelerated orbits which involve rotation in addition to
acceleration, event horizons exist as long as $a>\omega$. In the same 
way as discussed
for linear acceleration the field modes can be defined so that they 
are restricted to
one side of the horizons. This will involve a Bogoliubov transformation of the
Minkowski space creation and annihilation operators. The transformed 
vacuum state is
then the Fulling-Rindler vacuum associated with uniform acceleration 
of the $x'$
component of the space time trajectory. However, the Fulling vacuum 
{\em is not a
true vacuum state} since there now are negative energy states present 
in the right
Rindler wedge. Thus, the negative energy states are no longer restricted to the
space-time region behind the horizons, as it is for linear acceleration. On the
other hand these states are mainly located behind the static limit, 
but this region
is not causally disconnected from the reference curve
$C$. The presence of negative energy states implies that the 
Hamiltonian does not have a well
defined ground state even after separating the modes behind the 
horizons from those in front
of the horizons.

Even if the Fulling vacuum is no longer a true vacuum it is 
convenient to refer to
this state and the corresponding excitations in the description of 
the (generalized)
Unruh effect, since the modes that are causally disconnected are 
explicitly removed
from the description. As already discussed there are now two sources 
for excitation
of an accelerated detector. The detector can be excited by absorbing 
positive energy
excitations (which are present when the Minkowski vacuum state is 
expressed in terms of
Rindler space excitations) and it can be excited by emitting negative energy
excitations. This situation is quite analogous to the situation of a 
stationary
detector close to the static limit of a rotating black hole. In that 
case the latter
process, emission of negative energy excitations has a classical 
analogy in the {\em
Penrose effect} \cite{Penrose69}, where a physical system can gain energy by placing an object
into one of the negative energy orbits between the static limit and the
horizon.

We now turn to the case $\omega>a$, which we will discuss more 
briefly. The Hamiltonian
of the accelerated frame gets the simple form \pref{transh3}, which 
is composed by the
Minkowski space Hamiltonian $H$ and the rotation generator $J_z$. 
Since these two
operators commute, due to rotational invariance, the eigenstates of 
the accelerated
Hamiltonian $H_O$ are the same as those of the Minkowski space 
Hamiltonian, when these
are expressed as angular momentum states. The energy eigenvalues have the form
\be{rotenergy}
\e=\frac{\omega}{\sqrt{\omega^2-a^2}}\;E-\sqrt{\omega^2-a^2}\;m_z
\ee
where $E$ is the eigenvalue of $H$ and $m_z$ the eigenvalue of $J_z$.

In this case there are negative energy solutions when $m_z$ is 
sufficiently large compared to $E$. This means that Minkowski space 
is no longer the
ground state of the Hamiltonian and there is no other true vacuum state.  For
$a<\omega$ there is no event horizon, as we have already discussed, 
but there is a
static limit, and again one can show that the negative energy solutions have an
oscillatory behaviour outside the static limit, but are exponentially 
damped inside.
Related to the disappearance of the event horizons there is a lack of 
symmetry between
positive and negative eigenvalues in the spectrum of
$H_O$ \pref{rotenergy}. Thus, there is no longer a freedom of mixing 
creation and
annihilation operators    when the Hamiltonian is diagonalized. In 
this sense we are
stuck with the standard creation and annihilation operators. This means that
Minkowski space contains no excitations (relative to these operators) and the
excitation of a detector that is accelerated in a circular orbit is 
therefore (as
described in the accelerated frame) only due to the emission of negative energy
quanta.

We will however again stress that the qualitatively different 
descriptions of the cases
$a>\omega$ and $a<\omega$ is due to the difference of asymptotic 
behaviour of the
accelerated orbits. In reality the excitation (and de-excitation) of 
an accelerated
detector reaches equilibrium in a finite time determined by $a$ and 
$\omega$, and the
asymptotic form of the curves is therefore less important. This means that the
excitation spectrum changes continuously when $\omega$ is changed 
through the value
$\omega =a$, and we will show this explicitly in the next section.

\section{Particle detectors and thermodynamics}
For linear acceleration the Minkowski vacuum state has the form of a
thermally excited state of temperature $T_a= a/(2\pi k_B)$, when the 
state is expressed
in terms of the Fulling field quanta. This representation of the Minkowski
vacuum, as a hot state, has relevance also for (planar) accelerated 
motion {\em with
rotation}, when $a>\omega$ and therefore event horizons exist.

To examine the interpretation of the Minkowski vacuum in the 
accelerated frame when
$\omega\neq 0$, we return to the Bogoliubov transformation \pref{Bog} 
that relates the
Minkowski and Fulling field quanta. The corresponding transformation 
between the
vacuum states is
\be{Bogvac}
\ket{0_M}={\cal K} e^{-\suml_i \tanh\chi_i\, \tilde a_i^{\dag}\tilde 
b_i^{\dag}}\ket{0_F}\;
,\;\; {\cal K}=e^{-\suml_i \ln(\cosh \xi_i)}
\ee
where ${\cal K}$ is a normalization factor and  $\ket{0_M}$ is the
Minkowski vacuum state.
$\ket{0_F}$ is the Fulling vacuum state, which is the state annihilated by the
operators $\tilde a_i$ of the right Rindler wedge as well as the 
operators $\tilde b_i$
of the left Rindler wedge. The corresponding field modes we again label by
the index $i$. With the parameter
$\chi_i$ specified by
\pref{chsh} the Minkowski vacuum state can be written as
\be{Bogvac2}
\ket{0_M}&=& {\cal K}e^{-\suml_i e^{-\Omega_i\pi}\, \tilde a_i^{\dag}\tilde
b_i^{\dag}}\ket{0_F}\nn
&=& {\cal K}\prodl_i \suml_{n_i=0}^{\infty}
\frac{(- 1)^{n_i}}{n_i!}
\,e^{-n_i\Omega_i\pi}\;
\tilde a_i^{\dag n_i}\tilde b_i^{\dag n_i}
\ket{0_F} \nn
&=& {\cal K}\prodl_i \suml_{n_i=0}^{\infty}
(- 1)^{n_i}\,e^{-n_i\Omega_i\pi}
\ket{{n_i}_{\;L}}\otimes\ket{{n_i}_{\;R}}
\ee
where $n_i$ is the occupation number of the $i$'th mode, either in 
the right Rindler
wedge, denoted by $R$, or in the left Rindler wedge, denoted by $L$. 
This expression
for $\ket{0_M}$ is valid for linear as well as for non-linear 
acceleration ($\omega\neq
0$), since it is expressed in terms of the eigenvalues
$-\Omega_i$ of the boost operator rather than the energy of the field 
modes. The sum is
only over the {\em positive} values of
$\Omega_i$.

When the Minkowski vacuum state is expressed only in terms of the
field modes of the right Rindler wedge, which are the relevant ones 
for a detector
following the reference orbit
$C$, it takes the form of a mixed state rather than a pure state. In 
the expression above
this follows from the form of $\ket{0_M}$ as an {\em entangled state} 
between the
subsystems associated with the left and right Rindler wedges.  The 
corresponding reduced
density operator, where the states of the left Rindler wedge are 
traced out, has the form
\be{reddensop}
\rho_R&=&Tr_L\Big[\ket{0_M}\bra{0_M}\Big]\nn
&=& \prodl_i \suml_{n_i=0}^{\infty}
\,(1-e^{-2\pi\Omega_i})e^{-2\pi\Omega_i n_i}
\ket{{n_i}_{\;R}}\bra{n_{i\;R}}
\ee
For linear acceleration, with $\Omega_i=\e_i/a$, where $\e_i$ is the energy
of the $i$'th mode, the density operator has the form of a thermal 
distribution over
energy eigenstates and is consistent with the Bose-Einstein distribution
discussed earlier in the paper.

Our main object is now to discuss how this thermodynamic 
interpretation is changed
when
$\omega\neq 0$ and therefore the relation between the boost 
eigenvalues and the energy
eigenvalues is changed. There are two different points of view we may take.

The first point of view is to describe the excitation probabilities 
for an accelerated
detector, in this case too, as seen in the Rindler coordinate frame 
and not in the rest
frame of the detector. Compared with linear acceleration the relation 
between $\Omega_i$
and the energies of the field quanta then is changed only by the substitution
\be{rho}
a\ra\frac{a^2-\omega^2}{a}\equiv \frac{\rho^2}{a} \ee
where the
last expression is the proper acceleration of the {\em projected} 
orbit, which is
restricted to the $(x',t')$-plane. Thus, in this picture the Minkowski
vacuum state is a thermal state, in the same way as for linear 
acceleration, and all
field excitations restricted to the right Rindler wedge have positive energy.
A detector accelerated along the orbit $C$ will be excited by the field quanta
which are present in this state, but the excitation will not be thermal since
the detector {\em moves} through the ``gas" of excitations. This motion is a
constant drift in the $y'$-direction. This description fits the point of view
of Letaw and Pfautsch who refer to the vacuum state of all planar orbits with
$a>\omega$ as being identical to the Fulling vacuum
\cite{ref:LetawStationaryDetectorExcitation}.

The second point of view is to describe the Minkowski vacuum state in 
the accelerated
frame, which at all times is the rest frame of the stationary orbit $C$.
When restricted to the space-time region outside the horizon, this state
is also now described by the reduced density matrix \pref{reddensop}, but
the  relation between
the boost eigenvalues $\Omega_i$ and the energies $\e_i$ of the 
Hamiltonian $H_O$ is
\be{energyrot2}
\Omega_i=\frac{1}{\rho}\e_i-\frac{\omega}{\rho^2}{k_{y'}}_i
\ee
where ${k_{y'}}_i$ is the eigenvalue of the translation operator 
$P_{y'}$ in the the mode $i$.
Expressed in terms of the variables of the accelerated frame the 
density operator of
the Minkowski vacuum state is
\be{reddensop2}
\rho_R&=&|{\cal K}|^2 \prodl_i \suml_{n_i=0}^{\infty}
\,e^{-\frac{2\pi}{\rho}
n_i(\e_i-\frac{\omega}{\rho}{k_{y'_i})}}
\ket{{n_i}_{\;R}}\bra{n_{i\;R}}\nn
&=&|{\cal K}|^2  e^{-\frac{2\pi}{\rho}
(H_O-\frac{\omega}{\rho}{P_{y'})}}
\ee
where the operators $H_O$ and $P_{y'}$ are restricted to the space of 
states defined
outside the horizons (on the right Rindler wedge).

The reduced density operator may be interpreted as
the partition function for a statistical ensemble, where the exponent 
of the expression
\pref{reddensop2} specify the thermodynamic potential. One notes that 
the potential
gets contribution not only from the energy but also from the 
conserved momentum in the
$y'$-direction.
However, $P_{y'}$ is not the momentum operator in an instantaneous 
rest frame of the
accelerated orbit. Therefore it introduces translations not only in 
the space direction of
the accelerated frame, but also in the time direction. For this 
reason it may be natural
to replace it by another symmetry operator ${\cal P}$ of the 
accelerated frame, defined by
\be{newmom}
P_{y'}=\frac{a}{\rho}({\cal P}-\frac{\omega}{a}H_O)
\ee
This is a transformed symmetry operator related to the momentum 
operator in the inertial
rest frame (at $t=\tau=0$). Note however that the transformation 
\pref{newmom} is not
identical to a Lorentz transformation to the rest frame, since the accelerated
Hamiltonian $H_O$ rather than the Hamiltonian $H$ of the inertial 
rest frame enters in the
expression. At the position of the accelerated observer ${\cal P}$ coincides
with the translation operator $P_y$ of the rest frame, but $P_y$ is 
not a symmetry
operator, and the full expression involves also generators for rotations and
boosts,
\be{symop}
{\cal P} = P_y- \omega K_x -\frac{\omega^2}{a}J_z
\ee

Expressed in terms of the symmetry operator ${\cal P}$ the density 
operator has the form
\be{reddensop3}
\rho_R=|{\cal K}|^2  e^{-\frac{2\pi a^2}{\rho^3}
(H_O-\frac{\omega}{a}{\cal P)}}
\ee
This we may read as defining a thermodynamic state of a ``gas" with a 
non-vanishing
temperature and drift velocity. The coefficient of the
thermodynamic potential determines the (vacuum) temperature $T_{a\omega}$ as,
\be{temp}
k_B T_{a\omega}=\frac{\rho^3}{2\pi a^2 }=\frac{a}{2\pi 
}(1-\frac{\omega}{a})^{\frac{3}{2}}
\ee
This temperature is not identical to the temperature $\rho^2/2\pi a$ 
of the (vacuum) gas
as measured in the Rindler coordinate system (\ie on the projected, 
linearly accelerated
orbit). There is a factor $\frac{a}{\rho}$, which we interpret
as a time dilatation factor. This factor is introduced by the Lorentz 
transformation
between the rest frame of the gas and the rest frame of the accelerated orbit.

The coefficient of the  conserved momentum $\cal P$ represents the 
velocity of the
vacuum state relative to the detector (in
Ref.\cite{ref:GerlachThermalAmbience} referred to as a ``chemical potential"),
\be{vel}
v_{a\omega}=\frac{\omega}{a}
\ee
One should however note that the physical interpretation of this term 
is somewhat
ambiguous, since ${\cal P}$ is not a pure translation operator, but 
involves also
rotation and boost. Thus, in the rest frame the motion of the ``hot 
vacuum gas" is not
simply a linear drift \cite{footnote1}.

It is of interest to note that even if the two operators $H_O$ and 
${\cal P}$ have spectra that are unbounded from below, the specific combination that 
appears in the density operator has a spectrum that {\em is} bounded from below. 
Thus, the coefficient of ${\cal P}$ is not a parameter that can be changed arbitrarily, since only for the value given in \pref{vel} will the density operator be normalizable and therefore give a
well-defined thermodynamic state. One way to view this is that when the a drift velocity is introduced for the thermal ``vacuum gas" , in terms of  a non-vanishing value of $\omega$, this will change the thermodynamical potential not only explicitly through a change in the coefficient of ${\cal P}$, but also indirectly through the change in the Hamiltonian  $H_O$ \cite{footnote2}. 
The prefactor that determines the 
temperature, on
the other hand, can be changed without a similar consistency problem. 
In particular, if $T_{a\omega}\ra0$ without changing the operator
$H_O-v_{a\omega}{\cal P}$ (\ie neglecting the true coupling between these
variables), the thermodynamic state
\pref{reddensop3} will continuously change from the Minkowski vacuum state to the Fulling vacuum state restricted to the
right Rindler wedge.

We will now relate the discussion of the vacuum state to the response 
of an accelerated
detector. To this end, we use a simple de~Witt-type
detector (see e.g. \cite{ref:BirrellDaviesQFCurvedSpace}), which has 
the following
transition rate from an energy level ${\cal E}_m$ to another energy
level ${\cal E}_n$:

\begin{equation}
\Gamma_{mn}
= \mathcal{T}_{mn}
\int_{-\infty}^{\infty} \mathrm{d}\tau \, e^{-i\mathcal{E}_{nm} \tau}
\bra{0_M} \hat{\phi}(x(0)) \hat{\phi}(x(\tau)) \ket{0_M}
\label{transrate}
\end{equation}
where $\mathcal{T}_{mn}$ is the squared matrix element for the 
transition between the two
levels, $\mathcal{E}_{nm}\equiv\mathcal{E}_{n}-\mathcal{E}_{m}$ is 
the energy difference
between the final and initial state and
$x(\tau)$ is the space-time position of the detector at proper time 
$\tau$. In the above
expression we have used time translation invariance of the Green's 
function, $G^+(T,
T+\tau)
\equiv
\bra{0_M}
\hat{\phi}(x(T)) \hat{\phi}(x(T+\tau)) \ket{0_M} = \bra{0_M}
\hat{\phi}(x(0)) \hat{\phi}(x(\tau)) \ket{0_M} \equiv G^+(\tau)$, due
to the stationarity of the trajectory.

Since the region behind the horizons (in the left
Rindler wedge) is causally disconnected, only the component of the 
field operator restricted
to the right Rindler wedge will contribute. This component has the form
\be{Rfield}
\phi_R(x(\tau))=\suml_i \left [ e^{-i\e_i\tau}
\tilde f_i(\frac{a}{\rho^2})\;\tilde a_i+e^{i\e_i\tau}
\tilde f_i^*(\frac{a}{\rho^2})\;\tilde a_i^{\dag} \right ]
\ee
In this expression $\e_i$ is the field energy as measured in the rest 
frame of the detector
and the function $\tilde f_i(\xi')$ is the $\xi'$-dependent part of 
the stationary field
solution, given by the modified Bessel function in 
Eq.\pref{f-function}. (The detector is
located at position
$\xi'=a/\rho^2$.) For the correlation function of the field along the 
accelerated
trajectory this gives
\be{corrfunc}
\bra{0_M} \hat{\phi}(x(0)) \hat{\phi}(x(\tau)) \ket{0_M}
&=&\suml_i |f_i(\frac{a}{\rho^2})|^2\Big[e^{-i\e_i\tau}\mean{\tilde 
a_i^{\dag}\tilde a_i}_M  +
e^{i\e_i\tau}\mean{\tilde
a_i\tilde a_i^{\dag}}_M\Big]\nn
&=& 
\suml_i|f_i(\frac{a}{\rho^2})|^2\Big[e^{-i\e_i\tau}\frac{1}{e^{2\pi\Omega_i}-1}
+ e^{i\e_i\tau}\frac{1}{1-e^{-2\pi\Omega_i}} \Big]
\ee
where in the last line we have made use of the expression 
\pref{vacexcit} for the expectation
value of the product of an annihilation and a creation operator (with 
$\e_i/a$ replaced by
$\Omega_i$). The corresponding expression for the transition rate is
\be{transrate2}
\Gamma_{mn}
&=& 2\pi\mathcal{T}_{mn}
\suml_i
|f_i(\frac{a}{\rho^2})|^2(1-e^{-2\pi\Omega_i})^{-1}\Big[\delta(\mathcal{E}_{nm}-\e_i)
e^{-2\pi\Omega_i}  + \delta(\mathcal{E}_{nm}+\e_i) \Big]
\ee
The term proportional to $\delta(\mathcal{E}_{nm}-\e_i)$ corresponds 
to absorption of a
field quantum of energy $\e_i$ while the term proportional to 
$\delta(\mathcal{E}_{nm}+\e_i)$
is an emission term. For linear acceleration, with 
$\e_i=a\Omega_i>0$, transition up in
energy only gets a contribution from the absorption term and transition 
down in energy only
gets a contribution from the emission term. The ratio between these two 
rates then is simply
given by the factor $\exp(-2\pi\Omega_i)=\exp(-2\pi{\cal E}_{nm}/a)$, 
and for an equilibrium
situation this gives the ratio between the probabilities  of 
occupation of the two levels.
This ratio has the form of a Boltzmann factor corresponding to the 
Unruh temperature \pref{TU}.

When $\omega\neq 0$ there are two changes. Since 
$\Omega_i=\e_i/\rho-\omega k_{y'}/\rho^2$ is
no longer fixed by the energy difference between the two levels, the term
$\exp(-2\pi\Omega_i)$ cannot be factorized out in the ratio between 
transitions up and
down. This effect is due to the drift of the gas of
field excitations, which means that the probability distribution over 
excited states is no
longer determined only by the energy. The second effect is due to the 
fact that $\e_i$ can
take negative as well as positive values. Thus, transition up in 
energy gets a contribution
from the emission term as well as the absorption term, and this is the 
case also for
transition down in energy. Both these effects tend to make the ratio 
between transitions up
and down more complicated. It cannot simply be written as a Boltzmann 
factor. Thus, the
simple thermal probability distribution over energy levels of the 
detector that is found for
linear acceleration is no longer there for the more general stationary curves.

Even if a simple Boltzmann factor cannot be extracted in the general 
case, an effective
temperature can be defined by the ratio between the rates for 
transitions up and down (in
energy) between a pair of energy levels \cite{ref:BellLeinaas1}, \cite{Unruh99}. This
will be an {\em energy dependent} temperature, \ie it will not only depend on
the parameters $a$ and
$\omega$ of the accelerated trajectory, but also the energy difference $\Delta
\mathcal{E}$ between the two levels of the detector. With $\Gamma_+$ as the
rate for transitions up and
$\Gamma_-$ the rate for transitions down, the effective temperature is given by
\be{efftemp}
k_B T_{eff}(\Delta \mathcal{E})=\Delta 
\mathcal{E}\ln(\frac{\Gamma_-}{\Gamma_+})
\ee

\begin{figure}
\includegraphics[width=7cm]{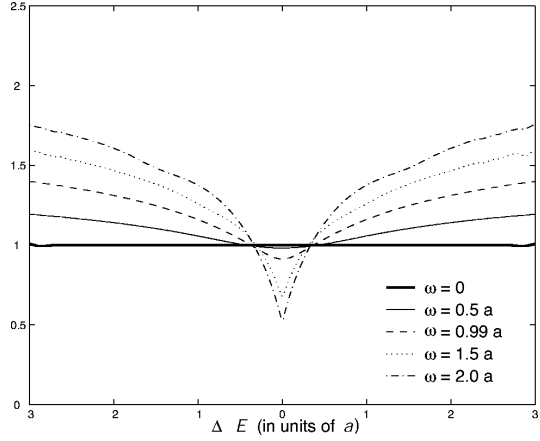}
\caption{The energy dependent effective temperatures.  The 
temperature is plotted as a
function of detector energy for several values of $\omega/a$. It is 
measured relative
to the Unruh temperature $a/2\pi$ defined by the proper acceleration 
of the accelerated
detector.}
\label{effectivetemps}
\end{figure}
In Fig.\ref{effectivetemps} we have plotted this function, relative to the
Unruh temperature
$a/2\pi$, for some values of $\omega/a$ both with $\omega<a$ and 
$\omega>a$. For
moderate values of $\omega/a$ one notes that the energy dependence is 
not very strong,
and this gives support to the idea that the definition of an 
effective temperature is
physically  meaningful. Even when $\omega/a$ is larger and increases 
beyond the value $1$
where the event horizon of the accelerated orbit disappears, there is no 
dramatic change.
Thus, even if in the general case a {\em precise} temperature cannot 
be given, a
qualitative description of the effect as {\em heating} of the 
accelerated detector by
the vacuum fluctuations seems reasonable, with an effective 
temperature not very different
from the Unruh temperature
$a/2\pi$ defined by the acceleration parameter of the orbit.

\section {Non-planar motion}
We now turn to the general case where the motion is no longer planar 
and the general form
of the Hamiltonian is given by \pref{accham} or \pref{accham2}, with 
$\bs a\cdot \bs
\omega\neq 0$. Since this case is qualitatively not so different from 
the case of planar
motion, we will discuss it more briefly. In this case as well, we can simplify the form of the Hamiltonian by a Lorentz transformation. It is now done 
by transforming the
coefficients of the boost operator and the angular momentum to 
co-linear form. The
transformation is, as for planar motion, a boost in the direction 
orthogonal to both $\bs
a$ and
$\bs\omega$ (the y-direction).

The transformation of the boost operator and the angular momentum 
under a Lorentz
transformation is
\be{Lorentz}
\bs K&=&\gamma (\bs K'-\bs\beta \times \bs
J'-\frac{\gamma}{\gamma +1}\bs\beta\; 
\bs\beta\cdot \bs K')  \nn
\bs J&=&\gamma (\bs J'+\bs\beta \times \bs K'-\frac{\gamma}{\gamma
+1}\bs\beta\; \bs\beta\cdot
\bs J')
\ee
which gives the following expression for the Hamiltonian in terms of 
operators of the
transformed inertial frame,
\be{KJtrans}
H_O=\gamma[H'+\bs\beta\cdot\bs P'-(\bs a
-\bs\beta\times\bs\omega)\cdot\bs K' -(\bs\omega+\bs\beta\times \bs
a)\cdot \bs J']
\ee
for $\bs\beta$ orthogonal to $\bs a$ and $\bs\omega$.
The coefficients of the two operators become co-linear if
\be{colin}
\bs\beta=\frac{\bs\omega\times\bs a}{2(\bs\omega\times\bs 
a)^2}\big[a^2+\omega^2-
\sqrt{(a^2-\omega^2)^2+4(\bs\omega\cdot \bs a)^2}\big]
\ee
We note that by a further transformation in the form of a shift in 
the location of the
spatial origin, the operator
$H'$ can be absorbed in $\bs K'$ and $\bs P'$ can be absorbed in $\bs J'$.

We specify the choice of axes for the original frame in the same way as given
by \pref{accham2}, with
$\bs a$ directed (with value $a$) in the $x$-direction and with 
$\bs\omega$ as a
vector with components
$\omega_x$ and $\omega_z$ in the $x,z$-plane. For the transformed 
frame we choose the
$x'$-coordinate to be in the direction of the co-linear boost and 
axis of rotation, as
specified by \pref{KJtrans} and \pref{colin}. This direction is 
rotated relative to the
$x$-axis in the $x,z$-plane. With these choices the Hamiltonian of 
the accelerated frame
gets the form
\be{Htrans}
H_O=\gamma[-\sqrt{(a-\beta\omega_z)^2+\beta^2\omega_x^2}\; K_{x'}
\pm\sqrt{(\beta a-\omega_z)^2+\omega_x^2}\;J_{x'}]
\ee
where $\beta$ is the component of $\bs\beta$ in the direction of 
$\bs\omega\times\bs
a$ as given by \pref{colin}, $\gamma=(1-\beta^2)^{-1/2}$ and the relative
signs of the two terms is determined by the sign of $\bs\omega\cdot\bs
a$.

This form of the accelerated Hamiltonian is not very different from
the one given by \pref{transh2} for planar acceleration with
$a>\omega$.  The main difference is that the translation operator
$P_{y'}$ is replaced by the angular momentum operator $J_{x'}$. This
implies that the accelerated trajectory (of the reference curve $C$),
as seen in the transformed frame, now is a superposition of a rotation
about the $x'$-axis and uniform acceleration in the
$x'$-direction. It is an accelerated screw-like motion.

In the same way as for planar acceleration, the asymptotic form of 
the trajectory is
dominated by the $x'$-component of the motion, and the location of 
the event horizons
is determined by the uniform acceleration of the $x'$-projection of 
the reference curve.
The location of the event horizons and static limit in the 
accelerated frame can be
found in the same way as for planar motion, by transformation to the 
inertial rest frame
at time $\tau=0$. The main difference between the two cases is that 
there is no longer
translational invariance in the
$z$-direction, so that the corresponding surfaces are curved also in 
this direction. We
do not give explicit expressions for these surfaces, but note that 
the qualitative
picture is the same as before. Between the accelerated observer on 
the trajectory $C$ and
the event horizons there is a static limit, where the stationary curves of the
accelerated frame change from being time-like to being space-like. Note that in 
the present case
there is always an event horizon, without any restriction on the 
values of $a$ and
$\omega$.

Since $J_{x'}$ commutes with $K_{x'}$ the eigenstates of the 
accelerated Hamiltonian can
be chosen as eigenstates of the boost operator, in the same way as 
for planar motion. The
difference is that the $x',z'$-dependent factor of these states 
should be angular momentum
states and not plane waves. Thus, by a Bogoliubov transformation the 
field modes
that are located behind the horizons, as viewed from the accelerated 
observer, can be
separated out and energy eigenstates restricted to the right Rindler 
wedge of the
boost operator
$K_{x'}$ can be defined. Also now there will be negative energy 
states, and these are
up to exponentially decaying tails located between the horizon and 
the static limit. The
Minkowski vacuum state, when restricted to the right Rindler wedge, 
is described by a
density matrix which is of the form
\be{}
\rho_M={\cal N}e^{-\frac{1}{\theta}(H_O+\bs\Omega'\cdot\bs J')}
\ee
with $\cal N$ as a normalization factor, $\theta$ and $\bs\Omega'$ given
by
\be{theta}
\theta&=&\frac{\gamma}{2\pi}\sqrt{(a-\beta\omega_z)^2+\beta^2\omega_x^2}\nn
\bs\Omega'&=&\gamma (\bs\omega + \bs\beta\times \bs a)
\ee

When we again 
interpret the exponent
as a thermodynamic potential, the form suggests that in the 
accelerated frame the
Minkowski vacuum state can now be viewed as a rotating, hot "vacuum 
gas". However, the
operator
$\bs J'$ is a rotation operator in the Rindler coordinate frame and 
not in the rest
frame. If we change to operators of the inertial rest frame and parametrize the
density operator
$\rho_M$ of the Minkowski vacuum state in the same way as we did for 
planar motion, it gets
the form
\be{reddensop4}
\rho_M=|{\cal K}|^2  e^{-\frac{2\pi 
\gamma}{\sqrt{(a-\beta\omega_z)^2+\beta^2\omega_x^2}}
(H_O-\beta{\cal P)}}
\ee
where the symmetry operator ${\cal P}$ now is given by
\be{symop2}
{\cal P} = \frac{1}{\beta}(\bs\beta\cdot \bs P- 
(\bs\beta\times\bs\omega)\cdot \bs K
-[(1+\beta^2)\, \bs \omega+\bs\beta\times\bs a]\cdot\bs J)
\ee
We note that the expression for the symmetry operator in the planar 
case is recovered in
the limit $\bs a\cdot \bs\omega\rightarrow 0$ (when $a>\omega$).

\section {Concluding remarks}
We have in this paper examined the ``generalized Unruh effect", which refers to
observable vacuum effects in general stationary coordinate frames. 
Such a stationary frame
will be characterized both by acceleration and rotation with respect 
to an inertial
rest frame. In the main part of the paper we have focused on planar 
motion. For accelerated orbits, event horizons will then exist provided the proper 
acceleration $a$ dominates the angular velocity $\omega$ as measured in an inertial rest frame.

When event horizons exist, the situation is similar to that of uniform
linear acceleration.  By mixing in a certain way (Minkowski) creation 
and annihilation
operators in the form of a Bogoliubov transformation, excitations 
associated with the two
sides of the horizons decouple, and a field theory restricted to the 
``physical" side of
the horizons may be defined. Relative to these excitations Minkowski 
vacuum will contain
excitations and can be characterized by a thermodynamical potential 
which depends on a
vacuum temperature as well as a drift velocity of the vacuum.

Although Minkowski vacuum can be interpreted as a thermodynamic 
state, when viewed in the
accelerated frame, a stationary detector will not show an excitation 
spectrum which can
be expressed simply in terms of a Boltzmann factor. There are two 
reasons for this. The
first one is due to the drift (and rotation) of the (thermodynamic) 
vacuum state. This
will give rise to non-universal thermal effects like the ones seen by 
a detector moving
with large speed through a hot gas. The other effect is due to the 
presence of negative
energy quanta which are associated with the region behind the static 
limit in the
accelerated frame. Thus, the detector may be excited either due to emission
of negative energy quanta or by absorption of positive energy quanta.

For accelerations with $a<\omega$, the event horizons disappear and 
the motion can be
viewed as uniform circular motion. Although these cases seem 
qualitatively different from
the ones characterized by $a>\omega$, and although Minkowski space no 
longer can be viewed as a
(non-trivial) thermodynamic state, we have demonstrated the smooth 
transition of the
effective (energy dependent) temperature as measured by an 
accelerated detector, when the
parameters are continuously changed from one type of motion to the other.

In the case of stationary, non-planar orbits, which can be viewed as 
circular motion
imposed on uniform linear acceleration, event horizons will always exist, and
qualitatively the description of the (generalized) Unruh effect for 
this type of motion
will be similar to that of planar motion with $a>\omega$.

Finally we have pointed to the close relation between the vacuum 
effects for detectors
following general stationary curves in Minkowski space and for stationary
detectors close to the event horizons of massive rotating black 
holes. For planar
motion we in the appendix explicitly show how the metric of the
accelerated frame is identical to the limiting form of the Kerr 
metric for points
close to the equator of the rotating black hole when the mass of the hole tends
to infinity.
\subsection{Acknowledgements}
JML thanks the Miller Institute for Basic Research in Science for financial support and hospitality during his visit to UC Berkeley. Supports from the Research Council of Norway and the Fulbright Foundation are also acknowledged.

\appendix

\section{Appendix.  Connection between Kerr spacetime and trajectories with
$\omega \ne 0$}
\label{app:Kerr}

In this appendix we compare the sitation of a stationary observer close to the static
limit of a rotating (Kerr) black hole with that of an accelerated observer in
Minkowski space. We show that in the limit where the mass M of the black hole tends to
infinity, space-time near the observer becomes flat, and the space-time trajectory
becomes identical to a stationary curve in Minkowski space. We restrict the discussion
to the case where the observer is located close to the equator of the black hole. In
this case the space time curve is a planar stationary curve characterized by a
constant proper acceleration $\bs a$ and angular velocity $\bs\omega$, where
$\bs a \perp \bs\omega$ and $\bs\omega\neq 0$ and $a > \omega$. As we shall show, the
two parameters $a$ and
$\omega$ can be related to the parameters of the black hole and the position of the
observer. (In the appendix we use $G=1$ for the gravitationol constant (rather than
$\hbar=1$) and still use $c=1$ for the speed of light.) 

At the equator of a rotating black hole
($\theta = \pi/2$), the Kerr metric expressed in Boyer-Lindquist
coordinates gets the following form:

\begin{multline}
\mathrm{d}s^2 = \left ( 1 - \frac{2M}{r} \right ) \mathrm{d}t^2 -
\frac{r^2}{r^2 + j^2 - 2Mr} \mathrm{d}r^2
- r^2 \mathrm{d}\theta^2 \\ \\ - \left ( r^2 + j^2 + \frac{2Mj^2}{r} \right ) 
\mathrm{d}\phi^2
+ \frac{4Mj}{r} \mathrm{d}t \mathrm{d}\phi
\label{eq:AppendixKerrMetric}
\end{multline}
where $M$ is the mass of the hole and $j$ is the angular momentum per unit mass of the
black hole, usually called $a$ in the literature, but renamed here to avoid confusion
with the acceleration parameter. In these coordinates the event horizon is
located at $r_h = M -\sqrt{M^2 - j^2}$ and the static limit at $r_s = 2M$.

We now want to show that for a space-time region centered around a static
trajectory with fixed coordinates $r > 2M, \theta = \pi/2, \phi = 0$
(arbitrarily chosen), there exists a mapping between the Boyer-Lindquist
coordinates of the black hole and a coordinate system of an accelerated
observer in Minkowski space. This mapping becomes an isometry (i.e. it maps the
Kerr metric into the metric of the observer following the stationary
trajectory) in the limit $M\rightarrow \infty$. 

Since we focus on space-time
points close to the static limit we write the radial coordinate as 
\be{rcoord}
r = 2M + q
\ee
with $q$ as a new coordinate to replace $r$. The $M \rightarrow \infty$ limit we
assume to be taken in such a way that $q/M \rightarrow 0$ and $j^2/M
\rightarrow 0$. The last condition is imposed in order for the distance between
the static limit and the event horizon to stay finite when $M$ tends to infinity.
With these approximations the metric outside the static limit is 
\be{ds2}
ds^2 = \frac{q}{2M}dt^2 -\frac{2M}{q+\frac{j^2}{2M}}dq^2 - 4M^2d\theta^2 - 4M^2
d\phi^2 + 2jdtd\phi
\ee 
and with a rescaling of the coordinates,
\be{rescale}
\bar t  =\frac{t}{2M},\quad \bar q  = 2Mq,\quad \bar z  = 2M\theta,\quad \bar y 
= 2M\phi
\ee 
it further gets the $M$ independent form
\be{metric}
ds^2 = \bar q {d\bar t}^2 -\frac{1}{\bar q+j^2}d\bar q^2-  d{\bar z^2} -  d\bar
y^2 + 2jd\bar td\bar y
\ee 

To compare this expression with the Minkowski space metric, as it appears in an
accelerated frame with $\omega\neq0$, we make use of the 
(accelerated) coordinates \pref{orbit}. In these coordinates the
accelerated observer is not stationary, but has a constant drift velocity
in the $y$-direction determined by the ratio $\omega/a$. To compensate for
the drift we redefine the $y$-coordinate
\begin{equation}
\tilde{y} = y' + \frac{\omega}{\sqrt{a^2 - \omega^2}} \, \tau'
\end{equation}
so that the accelerated observer is stationary in the coordinate system $(\tau'
;\xi' , \tilde y , z)$. As previously discussed, the values of the proper
acceleration $a$ and the angular velocity $\omega$ fix the $\xi'$ coordinate
to be $\xi'=a/(a^2-\omega^2)$, while the other space coordinates may be
chosen as $\tilde y =z=0$. We note that the spatial hyperplane defined by these
coordinates is not the observer's plane of simultaneity (as it is for the
accelerated coordinates $(\tau;\xi,\eta , z)$ previously used), since
the coordinates are not static. However, this is not a problem, since the same
thing is true of the Boyer-Lindquist coordinates.

In the coordinates $(\tau';\xi' , \tilde y , z)$, the Minkowski space metric
 takes the form
\begin{equation}
\diff s^2 = \left ( a^2 - \omega^2 \right ) \left ( {\xi'}^2 -
\frac{\omega^2}{\left ( a^2 - \omega^2 \right )^2} \right ) \, \diff
{\tau'}^2 - \diff {\xi'}^2 - \diff \tilde{y}^2 -dz^2-
\frac{2\omega}{\sqrt{a^2-\omega^2}} \, \diff \tilde{y} \, \diff
\tau'
\label{eq:AppendixMinkowskiMetric}
\end{equation}
This can be recast in a form similar to \pref{metric} by
redefining the coordinates. Thus, if we make the following identifications
between coordinates
\be{identific}
\bar q &=&\frac{1}{4}({\xi'}^2 - \frac{\omega^2}{\left ( a^2 - \omega^2 \right
)^2})\nn
\bar t&=&2\sqrt{a^2-\omega^2}\tau'\nn
\bar y&=&\tilde y\nn
\bar z &=&z
\ee
and if we identify $2j$ with the following function of $a$ and $\omega$,
\be{2j}
2j=\frac{\omega}{a^2-\omega^2}
\ee
then \pref{eq:AppendixMinkowskiMetric} reproduces exactly the metric \pref{metric}
derived from the black hole metric.

We note that the angular momentum parameter of the black hole,
$2j=\omega/(a^2-\omega^2)\equiv d_{hs}$, can also be interpreted as the distance
between the event horizon and the static limit, as measured in the instantaneous rest
frame of the stationary observer. This follows from the expressions \pref{horizons} and
\pref{statlim3} for the location of the horizon and static limit in the accelerated
frame. In the same way $a/(a^2-\omega^2)\equiv d_{ho}$ measures the distance from
the observer to the event horizon. Thus the dimensionless ratio $\omega/a$ corresponds
to the ratio between these two distances, and the limiting value
$\omega/a\rightarrow 1$ can be obtained by making $j$ large compared to the distance
between the static observer and the static limit.

In conclusion, we have shown how to map the coordinate system near
a stationary observer, located outside the static limit of a rotating black hole, to a
Minkowski space coordinate system where an accelerated observer is at rest, and to
show that in the limit $M\rightarrow\infty$ the mapping is an isometry. The two
parameters $a$ and $\omega$ of the accelerated orbit we have related to the angular
momentum $j$ of the black hole and the distance from the static observer to the event
horizon.

The mapping between these two seemingly different situations indicates that the
(generalized) Unruh effect associated with motion along stationary space-time curves
in Minkowski space is of similar form as the effect measured by a stationary detector
outside the stationary limit of a massive rotating black hole. However, a more detailed
discussion of this point depends on showing in what sense the Minkowski vacuum state
is the natural vacuum state of the black hole in the limit $M\rightarrow\infty$.


\begin{thebibliography}{99}
\bibitem{footnote1}
Even for linear acceleration one may 
note an ambiguity in
associating a specific temperature with the vacuum state as viewed in 
the accelerated
Rindler coordinates. The temperature $a/2\pi$ corresponds to the 
local temperature
measured at the position of the reference curve $C$, while at other 
points in the
accelerated frame the temperature will be different. This is similar 
to the situation in
a gravitational field, where the specification of a temperature of a 
hot gas will depend
on a choice of a reference point, since the locally measured 
temperature will vary in
space due to the redshift effect.
\end{thebibliography}
\end{document}